\documentclass{article}

\usepackage{PRIMEarxiv}

\usepackage[utf8]{inputenc} 
\usepackage[T1]{fontenc}    
\usepackage{hyperref}       
\usepackage{url}            
\usepackage{booktabs}       
\usepackage{amsfonts}       
\usepackage{nicefrac}       
\usepackage{microtype}      
\usepackage{lipsum}
\usepackage{fancyhdr}       
\usepackage{graphicx}       
\graphicspath{{media/}}     
\usepackage{cleveref}
\usepackage{booktabs} 
\usepackage{tabularx} 
\usepackage{hyperref}
\title{HH4AI: A Methodological Framework for AI Human Rights Impact Assessment under the EU AI Act
\thanks{\textit{\underline{Citation}}: 
\textbf{Authors. Title. Pages.... DOI:000000/11111.}} 
}

\author{
  Paolo Ceravolo\\
  Università degli Studi di Milano\\
  Milan, Italy\\
  \And
  Ernesto Damiani\\
  Università degli Studi di Milano\\
  Milan, Italy\\
  \And
  Maria Elisa D'amico \\
  Università degli Studi di Milano\\
  Milan, Italy\\
  \And
  Bianca de Teffé Erb\\
  Deloitte Financial Advisory S.r.l. S.B.\\
  Milan, Italy\\
  \And
  Simone Favaro\\
  Deloitte Financial Advisory S.r.l. S.B.\\
  Milan, Italy\\
  \And
  Nannerel Fiano \\
  Università degli Studi di Milano\\
  Milan, Italy\\
  \And
  Paolo Gambatesa\\
  Università degli Studi di Milano\\
  Milan, Italy\\
  \And
  Simone La Porta\\
  Deloitte Financial Advisory S.r.l. S.B.\\
  Milan, Italy\\
  \And
  Samira Maghool\\
  Università degli Studi di Milano\\
  Milan, Italy\\
  \And
  Lara Mauri\\
  Università degli Studi di Milano\\
  Milan, Italy\\
  \And
  Niccolò Panigada\\
  Università degli Studi di Milano\\
  Milan, Italy\\
  \And
  Lorenzo Maria Ratto Vaquer\\
  Deloitte Financial Advisory S.r.l. S.B.\\
  Milan, Italy\\
  \And
  Marta A. Tamborini\\
  Università degli Studi di Milano\\
  Milan, Italy\\
}

\begin{document}
\maketitle

\begin{abstract}
This paper introduces the HH4AI Methodology, a structured approach to assessing the impact of Artificial Intelligence (AI) systems on human rights, with a particular focus on ensuring compliance with the EU AI Act and addressing a range of technical, ethical and regulatory challenges. The paper emphasizes the transformative nature of AI, marked by autonomy, data-driven operations and goal-oriented design, and underscores how the EU AI Act provides a regulatory framework that promotes transparency, accountability and safety.

Key challenges include the definition and assessment of "high-risk" AI systems across industries and contexts, exacerbated by the lack of universally accepted frameworks and standards. The complexity and rapid evolution of AI systems further complicate compliance efforts. To address these challenges, the paper examines the relevance of standards and guidelines, particularly those from ISO/IEC and IEEE, that address critical issues such as risk management, data quality, bias mitigation, and governance.

Against this backdrop, the paper presents an innovative Fundamental Rights Impact Assessment (FRIA) methodology. This gate-based framework isolates relevant risks for in-depth assessment through phases that include an AI system overview, a human rights checklist, an impact assessment, and a final output phase. A filtering mechanism tailors the assessment to the specific characteristics of the system, focusing on impact areas such as accountability, AI literacy, data governance, and transparency.

Finally, the paper provides a fictional case study of an automated healthcare triage service to illustrate the structured application of the methodology. Through systematic filtering, comprehensive risk assessment, and mitigation planning, the FRIA methodology demonstrates how to effectively prioritize the most critical scenarios. It provides clear recommendations for remediation, thereby promoting greater alignment with human rights principles.
\end{abstract}


\section{Introduction}
\label{sec:introduction}

Artificial Intelligence (AI) refers to technologies that enable machines to perform tasks requiring human-like intelligence, such as reasoning, learning, decision making, and perception. The European Artificial Intelligence Act (EU AI Act) provides a detailed definition in Article 3(1), describing AI systems as machine-based technologies designed to operate with varying degrees of autonomy. These systems analyze inputs such as data to produce outputs - predictions, recommendations or decisions - that can influence physical or virtual environments. 

The EU AI Act fully captures the transformative nature of AI systems. The high degree of autonomy they can achieve allows them to operate, make decisions or take actions without continuous human intervention. They are also data-driven, relying on data - whether provided by humans or collected from their environment - to learn and adapt. This links their ability to achieve goals to the data collection and interpretation processes that these systems implement.

Notably, the AI Act does not limit its definition to a single class of technologies, but instead encompasses a wide range of methodologies. These include machine learning techniques (such as deep learning) as well as symbolic approaches such as logic-based reasoning and knowledge representation. This broad scope reflects the evolving and heterogeneous nature of AI, which adds to the challenge of assessing its risks and ensuring compliance.

The complexity of AI law extends beyond its definition. AI systems operate in highly dynamic environments, and their potential impact on fundamental rights depends not only on their technical characteristics, but also on the context in which they are deployed. As a result, determining whether an AI system is high-risk requires careful analysis of both technological and legal factors. Given these challenges, a structured methodology is essential to systematically assess AI systems and ensure that their risks are effectively identified, analyzed, and mitigated in accordance with legal and ethical principles.


The risks associated with AI systems vary across industries and depend on specific use cases, making a one-size-fits-all definition impractical. While the EU AI law sets out compliance requirements, it provides limited guidance on how to conduct these assessments in practice. The technological diversity of AI-which includes machine learning, robotics, and symbolic reasoning-requires tailored assessment methodologies that can account for the unique characteristics of each approach. In addition, AI is rapidly evolving, rendering static standards obsolete and complicating long-term compliance efforts.

Several factors contribute to the complexity of AI assessment:

\begin{itemize}
    \item The interdependence of models, data, and external factors creates a dynamic environment with unpredictable interactions.
    \item Continuous retraining and updates can change the behavior of an AI system over time, often without immediate visibility.
    \item The lack of universally accepted frameworks makes it difficult to establish consistent assessment practices.
    \item Different priorities across jurisdictions - from safety to fairness - lead to regulatory inconsistencies.
\end{itemize}

Ensuring fairness and transparency presents additional challenges. Detecting and mitigating bias requires specialized tools and expertise, especially for opaque models such as deep learning. Accountability remains a key concern, as many AI systems operate as "black boxes," making it difficult to track decision-making processes.

The global implications of the AI Act add another layer of complexity. Alignment of EU regulations with international frameworks is essential to avoid trade barriers and foster innovation, but this will require extensive cross-border cooperation. Finally, resource constraints - including the high cost of compliance and the shortage of AI ethics and evaluation experts - place a disproportionate burden on companies, especially small and medium-sized enterprises (SMEs).

Given these challenges, a structured methodology is critical to effectively navigate AI risk assessment, ensure compliance, and promote trustworthy AI.

\section{Legal and Regulatory Background}
\label{sec:legal-background}

\subsection{The Challenges of AI Assessment}
\label{subsec:challanges}

The EU AI Act \cite{AIAct} provides a comprehensive regulatory framework to govern the design, development and deployment of AI systems within the European Union. It introduces a broad definition of artificial intelligence (Article 3) that emphasizes the autonomy of systems, their data-driven and adaptive nature, and the potentially implicit goals they may pursue. It also refers to a wide range of methods and tools, including planning, reasoning, knowledge representation and learning, as highlighted in Recital 12.

The AI Act \cite{AIAct} also establishes guidelines for risk management procedures (Article 9). High-risk AI systems are required to implement a risk management system, while low-risk systems may adopt such measures voluntarily. The risk management process includes the identification of risks, the assessment of their potential impact, and the application of appropriate risk mitigation strategies.

The Act \cite{AIAct} also outlines data governance and reporting requirements. Compliance with the \textit{General Data Protection Regulation} (GDPR) is essential (Article 10), as is adherence to cybersecurity principles (Article 15) and data quality requirements (Articles 10 and 15)\cite{gdpr}. Systems must implement robust quality control mechanisms (Article 17), maintain detailed technical documentation (Article 11), and comprehensively log system activities (Article 12). High-risk systems must be registered in a public database to ensure transparency and accountability (Article 13). In addition, provisions are included to ensure both transparency and human oversight (Articles 13 and 14).

The AI Act mandates conformity assessment procedures for vendors. Internal conformity assessments (Articles 16 and 43) require thorough and documented verification of compliance with the Act's requirements. Independent bodies are involved in conformity assessments for biometric systems (Article 43). In addition, compliance with harmonized standards approved and published by the European Commission presumes compliance with the Act (Article 40) \cite{AIAct}.

To summarize, the AI Act defines several key requirements:
\begin{itemize}
    \item \textbf{Procedural Requirements:}
    \begin{itemize}
        \item Risk management methodologies.
        \item Control and documentation procedures.
        \item Conformity assessment processes.
    \end{itemize}
    \item \textbf{Technical Requirements:}
    \begin{itemize}
        \item Adherence to harmonized standards, codes of practice, or codes of conduct.
        \item Implementation of best practices and guidelines.
    \end{itemize}
\end{itemize}

Harmonized standards, codes of practice \cite{bartle2005self, fichter2013voluntary}, and codes of conduct \cite{antonucci2018codes} are envisioned as critical tools to assist organizations in demonstrating compliance with the requirements of the AI Act. These resources will provide structured technical specifications and practical guidance to ensure a consistent approach to risk management, data governance and transparency. However, these standards are not yet available and are not expected to be released for another 2-3 years.

This delay presents a significant challenge for organizations seeking immediate compliance, as the lack of harmonized standards leaves room for uncertainty in interpreting and implementing the Act's requirements. In the meantime, organizations must bridge this gap by leveraging existing frameworks. International standards, such as ISO/IEC 23894, offer a valuable starting point by providing foundational principles that align with the Act's objectives. In addition, best practices and guidelines developed by industry associations, research institutions, and other organizations can serve as interim references to ensure adherence to core compliance principles.

While these interim solutions can mitigate the challenges of the lack of harmonized standards, organizations should remain vigilant. Adopting flexible and adaptive approaches will be essential to ensure that current compliance strategies can be seamlessly updated once harmonized standards and codes of practice are officially released. This transitional period underscores the importance of fostering collaboration across industries and stakeholders to share insights and align practices in preparation for the upcoming standardized frameworks.

Despite its comprehensive framework, implementation of the AI Act presents several significant challenges. A universally accepted reference framework has yet to be established, leaving organizations without a clear foundation to guide their compliance efforts. Risk assessment remains highly contextual, varying based on the specific application rather than the underlying technology, making it difficult to replicate and ensure consistency in assessments.

Adding to this complexity, standards are constantly evolving to keep pace with methodological advances and emerging technologies, creating a moving target for compliance. The required level of detail for assessments is also unclear, particularly when balancing self-assessment with empirical validation. Finally, there is considerable uncertainty about which aspects of compliance can be determined in advance and which require ongoing, real-time monitoring.

\subsection{Human Rights and Ethical Considerations}
\label{subsec:human-rights}

At this point, it is necessary to underline the role of human rights and ethical principles in shaping AI assessment methodologies.
The use of artificial intelligence systems inevitably impacts the human rights protected both at the national level and at the European and supranational levels. This necessarily entails implications not only from an ethical and moral point of view but also legally.
The difficult objective that States and International Organizations are trying to achieve is to find the right balance between the free use of artificial intelligence and the protection of individuals' fundamental rights. The framework within which these entities operate is defined by the Constitutions of individual States (for civil law countries), the role of judges (for common law countries), the different Charters of rights, and other legal sources that regulate the matter. This creates a multi-level system of guarantees and protection that fully enables the defense of individuals' positions, which are expressed both in an individual and collective dimension. In fact, the issue does not only concern the position of individuals but also how they relate to each other as a community. In short, this is an approach that puts the person at the center: a human-centered AI.
In this way, we try to prevent the new challenges of law, including rapid technological development, from catching legal systems unprepared. On one hand, it is right to preserve market logic, but on the other hand, this logic must be governed by a principle of certainty in legal relations.
In this regard, the aforementioned AI Act constitutes a valid example of legislation, adopted at the European level, which moves in this direction. It is a regulation aimed at regulating the use of AI within the framework of rights enshrined, primarily, in the Charter of Fundamental Rights of the European Union. The Preamble states that the objective of this regulation is to improve the functioning of the internal market by laying down a uniform legal framework in particular for the development, the placing on the market, the putting into service and the use of AI systems in the Union, in accordance with Union values, to promote the uptake of human-centric and trustworthy artificial intelligence (AI) while ensuring a high level of protection of health, safety, fundamental rights as enshrined in the Charter of Fundamental Rights of the European Union (the ‘Charter’), including democracy, the rule of law and environmental protection, to protect against the harmful effects of AI systems in the Union, and to support innovation."
From the perspective of the impact on fundamental rights, those most at stake include: the principle of equality and non-discrimination, the right to privacy, the right to transparency of operations performed by artificial intelligence, and environmental protection. These are rights with a generalized scope, which are involved with every use of an AI system. Alongside these, other more specific rights can be highlighted, which manifest in more detailed situations and under certain circumstances. For example, consider how the right to education might be impacted by different levels of literacy and understanding of artificial intelligence tools.
This section does not aim to provide a complete overview of all human rights involved, but rather to highlight how, in light of some of these rights, the impact of artificial intelligence is evident and significant.
Focusing, therefore, on the rights mentioned above, it can be emphasized that in terms of equality and non-discrimination, a crucial role is played by those who design the AI system and those who train it. Indeed, during the creation and testing phase, it is central not to include instructions that carry biases. These risks could persist with every use of artificial intelligence, thus reinforcing discrimination rather than removing it. This risk is even greater since AI systems merely reproduce reality, which is currently characterized by various forms of discrimination and bias.
As for the right to privacy, it is of particular importance during the data processing phase by AI systems. These systems process a multitude of information that can lead to unlimited surveillance of people's activities, through, for example, tools that process personal and biometric data, posing a risk to the safety of individuals and States.
Regarding the right to transparency of AI operations, it is essential that all users know how AI works and what the logical steps are, as well as the sources from which the information is drawn. This principle is closely linked to the prohibition of discrimination and the right to privacy. As for the first, by knowing the various steps undertaken by AI, users can verify the accuracy of the information and the absence of bias. In other words, it ensures that decisions are made without prejudice. As for the second, it allows for a check on the non-use of personal information in the decisions made by artificial intelligence systems.
A final aspect not to be overlooked concerns environmental protection in terms of the principle of sustainable development and the protection of future generations. It is undeniable that AI systems, in order to function, require a constant supply of energy, which in turn causes overheating in the servers that manage them. Without going into technical details, it is enough to say that AI systems, however they may be used to promote environmental sustainability (e.g., by reorganizing work processes), can, if widely used, pose a serious risk to the planet's balance.
These final considerations are connected to the ethical and moral issues surrounding the use of AI. The questions are numerous and all revolve around the responsible use of artificial intelligence systems. The timeless debate on the relationship between machines and humans thus finds new life and a new chapter. Finding the right way to limit the autonomy of machines is crucial for safeguarding people's safety in numerous areas of life, such as work and health, but not only. The ability of these systems to learn and improve ever more quickly constitutes another problem that must be addressed, along with the issue of accountability for the actions of artificial intelligence.
In this context, ethical aspects and human rights intersect. It is, therefore, emphasized that the transparency of AI decisions and the continuous monitoring of these decisions are essential to avoid the creation of inequalities. Similarly, there must be the creation of a clear regulatory framework regarding responsibility for damages caused by artificial intelligence. Crucially, it is essential to create a legal system that is not only focused on safety and risk prevention but also on the central role of human control, based on shared education and awareness.

\subsection{International Frameworks}
\label{subsec:international-frameworks}
Discuss key legal and regulatory frameworks, such as the OECD AI Principles, EU AI Act, and UNESCO guidelines, emphasizing their relevance to AI assessment.

The European Union’s AI Act represents a binding regulatory approach that categorizes AI systems by risk level, imposing strict legal obligations on high-risk applications while prohibiting those deemed unacceptable. With its extraterritorial reach, the Act ensures that AI systems impacting the EU market adhere to transparency, human oversight, and accountability standards. In contrast, international soft-law frameworks, such as the OECD AI Principles, the UNESCO Recommendation on the Ethics of AI, and the Council of Europe’s Framework Convention on AI, Human Rights, Democracy, and the Rule of Law, establish voluntary guidelines emphasizing fundamental rights, fairness, and responsible AI governance. While these frameworks influence global policy, they rely on voluntary compliance and lack direct enforcement mechanisms. The United States, by contrast, follows a decentralized, sector-specific approach with no overarching federal AI law. Instead, it relies on a combination of existing statutes, agency guidance, and state-level regulations, with organizations like the National Institute of Standards and Technology (NIST) issuing non-binding frameworks such as the AI Risk Management Framework, which provides voluntary risk assessment principles for AI systems. The fragmented regulatory landscape in the U.S. creates inconsistencies and gaps, raising ongoing debates about the need for a more cohesive federal strategy. The divergence between the EU’s legally binding model and the U.S.’s market-driven, self-regulatory approach highlights broader tensions in global AI governance. While international organizations push for regulatory alignment through high-level principles, national laws reflect different priorities, creating challenges in achieving cross-border interoperability. The EU AI Act’s influence is already evident in discussions on AI regulations in Canada, Japan, and Brazil, which are exploring risk-based models. However, differences in enforcement strategies and legal traditions may limit global harmonization. As AI technologies evolve, the interplay between binding regulations, voluntary principles, and sector-specific laws will shape the future of AI governance, underscoring the need for continued international cooperation to manage AI’s risks and benefits effectively.

This overview sets the legal background for an AI impact assessment strategy based on an interdisciplinary approach and focusing on fundamental human rights. As set forth in previous paragraphs, the absence of an unambiguous and compact international regulatory framework makes it imperative for academics and practitioners to draw a roadmap to accompany companies in this new challenge.

\section{Standards and Guidelines}
\label{sec:standards}

\subsection{Standards for AI Assessment}
\label{subsec:ai-standards}
The assessment of AI systems relies on established standards and frameworks that provide guidance on risk management, transparency, and accountability. Among the most relevant are the ISO/IEC \cite{oviedo2024iso} and IEEE standards \cite{schiff2020ieee, winfield2021ieee} and frameworks developed by the National Institute of Standards and Technology (NIST) \cite{ai2024plan}. 

Of particular importance is ISO/IEC 23894 \cite{floridi2023brussels}, which addresses risk management for AI systems and provides a structured approach to identifying, assessing, and mitigating potential risks. This standard closely aligns with regulatory requirements such as those outlined in the AI Act, making it a valuable resource for organizations seeking to ensure compliance. In addition, ISO/IEC 25012 \cite{gualo2021data} focuses on data quality, emphasizing critical aspects such as accuracy, completeness, and consistency, which are essential for AI systems that rely heavily on high-quality data sets for training and operation. Another notable contribution is ISO/IEC TR 24027 \cite{iso2021iso}, which provides methods for identifying and mitigating bias in AI systems to ensure fairness in their use. Governance considerations are also addressed by ISO/IEC 38507 \cite{iso2022iec}, which provides guidance on integrating AI into organizational governance structures to enhance accountability and oversight.

To further complement the above standards, ISO/IEC 42001 \cite{iso42001} and ISO/IEC 42005 \cite{iso42005iec} offer additional frameworks designed specifically for managing AI systems throughout their lifecycle. ISO/IEC 42001 outlines the requirements for an AI management system, helping organizations establish processes for continual monitoring, evaluation, and improvement of AI system performance while ensuring alignment with ethical principles and regulatory demands. ISO/IEC 42005, which is currently under development, will focus on AI system impact assessments. Once finalized, it will provide a structured approach to evaluating the potential social, environmental, and economic impacts of AI systems. The standard is intended to apply to any organization involved in developing, providing, or using AI technologies. It will outline when and how to conduct these assessments, ensuring that the impact evaluation process is effectively integrated into existing AI risk management and management systems. Additionally, ISO/IEC 42005 will offer guidance on improving the documentation of impact assessments, helping organizations maintain transparency and accountability in their AI operations.

In parallel, NIST has developed frameworks that are equally important for assessing AI systems. The NIST AI Risk Management Framework (AI RMF) \cite{nist_airmf} serves as a flexible and voluntary guide for identifying, assessing, and managing AI-related risks. This framework emphasizes a comprehensive and iterative risk management process, making it particularly relevant for organizations seeking to address the dynamic challenges posed by AI technologies. NIST has also introduced the AI 600-1 \cite{nist_600} standard, which is particularly focused on addressing the unique risks associated with generative AI technologies. As generative AI systems continue to evolve and gain prominence, they introduce novel challenges—such as the potential for creating harmful content, bias in generated outputs, or misuse of generated data. The AI 600-1 standard helps organizations identify these specific risks and proposes actionable strategies for mitigating them. In addition, the NIST Privacy Framework, while not specific to AI, provides valuable insights into managing privacy risks, a critical concern for AI systems that process sensitive or personal data.

As part of the IEEE Global Initiative on Ethics of Autonomous and Intelligent Systems, which aims to promote the ethical use of technology and address the societal impacts of AI and other emerging technologies, the IEEE 7002-2022 \cite{olszewska2022ieee} standard provides guidelines for ensuring ethical practices in the development and deployment of AI systems. It focuses on aspects like accountability, transparency, fairness, and safety, and is intended to guide organizations in making responsible decisions when creating AI technologies. In conjunction with this, IEEE 7010-2020 \cite{schiff2020ieee} provides guidance for assessing the impact of AI systems on human well-being, ensuring that AI and autonomous systems are developed to promote human safety and rights, particularly in areas such as healthcare, transportation, and personal data. Together, these standards help align AI development with both ethical considerations and the promotion of human wellbeing.

These standards and frameworks collectively address fundamental challenges in AI assessment, such as risk management, data quality, bias mitigation, and governance. Although they provide a solid foundation for compliance and best practices, they fall short in providing the level of detail required for specific scenarios. In particular, they lack a precise set of procedures for assessing compliance with specific requirements identified in the AI Act \cite{AIAct}. Their applicability often depends on the context and use case of the AI system, requiring organizations to interpret and adapt these guidelines to their unique situations. As a result, organizations are encouraged to adopt a customized approach that combines multiple standards and frameworks to fill in the gaps and effectively address both technical and regulatory requirements.

\subsection{Guidelines from Research and Industry}
\label{subsec:guidelines}
Recent advances in AI have prompted leading institutions and stakeholders to establish influential guidelines for the assessment and evaluation of AI systems. The \textbf{Alan Turing Institute} has proposed a comprehensive framework that emphasizes the importance of \textit{transparency}, \textit{accountability}, and \textit{robustness} in AI systems, advocating rigorous testing against adversarial scenarios and the use of explainability tools such as SHAP (SHapley Additive exPlanations) and LIME (Local Interpretable Model-agnostic Explanations) to ensure interpretability \cite{leslie2019understanding}. Similarly, the \textbf{European Union Agency for Cybersecurity (ENISA)} has outlined guidelines focused on \textit{security} and \textit{resilience}, recommending continuous monitoring, risk assessment, and the adoption of standardized metrics to evaluate the performance of AI systems under varying conditions \cite{ntalampiras2023artificial}. Other stakeholders, including the \textbf{Partnership on AI}, have highlighted the need for \textit{fairness} and \textit{bias mitigation}, promoting the use of fairness-conscious algorithms and diverse datasets to reduce discriminatory outcomes \cite{heer2018partnership}.

\subsection{Overview of Tools}
\label{subsec:tools-overview}

As AI systems increasingly permeate critical domains, the risk of human rights violations arising from their misuse or misalignment with ethical principles becomes a concern. To address these challenges, several organizations have developed tools and frameworks aimed at assessing and mitigating AI-related risks. Below, we compare some noted tools, in order to specify their strength and learn from their limitations.

\textit{Microsoft’s Responsible AI Impact Assessment (RAIIA)} is a structured framework designed to ensure AI systems are developed and deployed responsibly. It provides organizations with templates and guidance to assess AI systems against key principles like fairness, reliability, transparency, privacy, and inclusiveness. 

\textit{Google} also has provided a toolkit that comprises a suite of tools and frameworks to assist developers in creating responsible AI systems. Key components include Explainable AI (XAI), fairness indicators, and model cards. 

AI Fairness 360 (AIF360) is an open-source toolkit by \textit{IBM} designed to detect and mitigate bias in machine learning models. It includes metrics, algorithms, and visualization tools for fairness.

In addition, OpenAI also provides guidelines and best practices for responsible AI use, focusing on applications of large language models and generative AI systems.

In efforts from research institutes, \textit{Ethical AI Toolkit by the Montreal AI Ethics Institute}, focuses on ethical dilemmas and includes worksheets for ethical impact assessments.  While this Toolkit takes a holistic approach, emphasizing on societal impact, it lacks technical depth and automation. 

\textit{Hugging Face’s Model Evaluation Tools} also provide insights into performance and fairness for pre-trained NLP models. It makes a huge effort in explanaibility of models but has a narrow applicability to specific model types.

\subsection{Comparison and Insights}
\label{subsec:tools-comparison}
By comparing the standards, guidelines, and tools, significant insights emerge regarding their ability to address the fundamental challenges of AI impact assessment. Each approach brings unique strengths, such as providing structured methodologies for assessing transparency, accountability, and ethical considerations. However, they also present limitations, including challenges related to scalability, adaptability to fast-evolving technologies, and inconsistencies in application across different sectors and contexts. Examining these strengths and weaknesses allows us to better understand how effectively these approaches support comprehensive AI impact evaluations and where further improvements are necessary, ultimately helping to identify a strategy that successfully integrates both regulatory and technical requirements.

\begin{table}[]
\centering
\footnotesize
\renewcommand{\arraystretch}{1.3} 
\begin{tabularx}{\textwidth}{|l|X|X|}
\hline
\textbf{Standard} & \textbf{Strengths} & \textbf{Limitations} \\
\hline
\textbf{ISO/IEC 23894} & Provides a framework for managing risks in AI systems. Focuses on safety, security, and ethical implications. Emphasizes continuous monitoring and improvement. & Narrow scope for broader AI applications. May not address all industry-specific needs. Less applicable in non-technical contexts.\\
\hline
\textbf{ISO/IEC 25012} & Defines data quality requirements critical for AI system evaluation. Provides criteria for ensuring data impacts are considered in AI evaluation. & Focuses more on data quality than AI-specific impacts. Lacks direct guidelines for ethical or societal implications. \\
\hline
\textbf{ISO/IEC TR 24027} & Provides guidelines for AI explainability and transparency. Focuses on interpretability, traceability. Ensures that AI outputs are understandable to stakeholders. & High-level approach with limited actionable guidance for specific use cases. Challenging to apply in highly specialized technical fields. Requires significant customization for specific industries. \\
\hline
\textbf{ISO/IEC 38507} & Provides governance principles for AI, supporting responsible AI decision-making. Strong focus on accountability, fairness, and legal considerations. & May be too general, lacking specific metrics for AI impact assessment. Implementation can be complex for organizations lacking governance frameworks. \\
\hline
\textbf{ISO/IEC 42001} & Guides organizations on how to implement ethical AI practices in their processes. Supports ongoing evaluation and adjustment of AI systems' impact over time. & Doesn't provide detailed metrics or concrete tools for impact measurement. Difficult to implement for organizations without AI expertise. \\
\hline
\textbf{ISO/IEC DIS 42005} & Aims at defining ethical frameworks for AI, ensuring sustainable development. Encourages responsible AI lifecycle management. & Not finalized (as a draft), making its adoption and practical application challenging. May be complex to implement fully across all sectors. \\
\hline
\textbf{NIST AI RMF} & Provides a comprehensive, risk-based framework for assessing AI systems' impacts. Covers governance, transparency, and performance evaluation. & Complex and may require significant resources and time to implement in full. Lacks prescriptive detail in terms of actionable, operational steps. \\
\hline
\textbf{NIST AI 600-1} & Focuses on AI-specific security and privacy guidelines. Provides risk assessment models specific to AI. & Primarily security-oriented, with limited focus on broader social, ethical, or human rights implications. Risk assessment may not fully address non-technical concerns like bias. \\
\hline
\textbf{IEEE 7002-2022} & Strong focus on ethical design and human-centric approaches. Provides clear guidelines for system-level AI ethics. & Primarily a guiding framework, without direct, measurable implementation steps. May require translation into actionable regulatory frameworks. \\
\hline
\textbf{IEEE 7010-2020} & Provides guidelines for assessing and mitigating the societal impacts of AI, especially in terms of well-being. Focuses on long-term sustainability and human-centered design. & Limited scope in addressing broader societal and environmental concerns. Might be difficult to quantify or apply consistently across all AI domains. \\
\hline
\end{tabularx}
\caption{Comparison of AI-related standards: key strengths and limitations.} 
\label{tab:std}
\end{table}

By putting the lens on the standards, it becomes clear that they highlight a diverse range of focuses and intended users, with some prioritizing technical evaluation, others addressing governance or fairness, and a few focusing directly on societal impacts. For instance, ISO/IEC 23894 and ISO/IEC 25012 focus on system evaluation and data quality, respectively, but lack a broader risk management perspective and do not provide guidance on long-term risks. ISO/IEC TR 24027 serves a crucial role in creating a shared understanding of AI terminology, fostering clearer communication and better alignment across stakeholders, yet it falls short in providing actionable steps for mitigating AI risks. 

ISO/IEC 38507 has a governance-centric focus that may not sufficiently address the technical risks associated with AI systems. At the same time, ISO/IEC 42001 emphasizes governance and strategic oversight, making it ideal for organizations seeking strong AI oversight, but it may not fully address technical risks and could be challenging to integrate with existing frameworks. ISO/IEC DIS 42005 provides a comprehensive lifecycle framework for continuous monitoring and accountability, fostering a holistic approach to AI risk management. However, being in draft status and having a broad scope, it may be complex to implement, especially for smaller organizations.

IEEE 7002-2022 and IEEE 7010-2020 focus on the ethical and social impacts of AI, making them essential for addressing privacy and societal harm, but they may overlook broader technical and operational risks. In contrast, the NIST AI RMF presents a detailed and practical approach to risk management across different sectors, but it may be challenging and resource-demanding for smaller organizations to implement, while NIST AI 600-1 provides targeted guidance for critical infrastructure, though it is highly technical and may not be applicable to non-infrastructure sectors.

Table \ref{tab:std} presents an overview of the key strengths and limitations of each analyzed standard.

We now shift our focus to the tools, assessing their core strengths and drawbacks. Firstly, RAIIA offers comprehensive guidance, detailed questionnaires and prompts that help identify potential risks across the AI lifecycle. This tool is built around Microsoft’s Responsible AI principles, ensuring consistency with widely accepted ethical standards. It also fosters accountability by encouraging collaboration among diverse stakeholders and benefits from scalability for use across industries and AI applications. Several limitations of this tool are resource intensiveness, lack of automation and, its	proprietary nature since is heavily tied to Microsoft’s ecosystem, potentially limiting accessibility for organizations using other platforms.

From the strength points of Google’s AI Toolkit we can mention, Explainability Features that offer tools for interpretation and visualization of models' behavior, enhancing transparency and trust. It also provides metrics and visualization tools to identify biases in datasets and models. This toolkit encourages documentation of model characteristics, promoting accountability and informed usage by model cards. In this kit several tools are available as open-source projects, increasing accessibility and adaptability. 

Google's AI Toolkit, focuses highly on technical aspects which means it primarily targets developers and may not address broader organizational or societal concerns. It also requires technical expertise to implement and interpret outputs effectively as it faces Steep Learning Curve at some points. It also lacks contextual guidance and tailored recommendations for industry-specific risks and mitigation strategies.

On the other hand IBM’s AI Fairness 360, provides Algorithmic Bias Mitigation in pre-processing, in-processing, and post-processing techniques to address bias. It also implements comprehensive Metrics measurements, over 70 fairness metrics, serving for various definitions of fairness. Its open-source platform is easily customizable and integrable into different workflows. Despite its valuable contribution, using tool is complex and requires advanced knowledge of machine learning. It has a narrow focus, concentrates on fairness, with limited tools for broader aspects like transparency and accountability. Moreover, it has not been optimized for deployment in large-scale, real-time systems.

OpenAI’s Use Case Guidelines provide context-specififc guides offering tailored recommendations for specific use cases, such as content generation or decision support. It conveys ethical considerations and emphasizes on avoiding harm and ensuring user safety. The tool contain educational resources which includes tutorials and documentation to improve user understanding whilst it lacks dedicated software or automated frameworks for risk assessment. It also provides high-level recommendations and conceptual guidance rather than detailed implementation strategies.

In a holistic view, we witness the following strengths across above-mentioned tools:

\begin{itemize}
    \item Promoting Transparency: Most tools include features like documentation (e.g., model cards) and explainability frameworks.
    \item Bias Detection: Many tools provide metrics and methods to identify and mitigate bias, addressing fairness concerns.
    \item Accessibility: Open-source options (e.g., IBM AIF360, Google’s tools) allow broad adoption and customization.
\end{itemize}

while still following limitations should be addressed:

\begin{itemize}
    \item Technical Barriers: Most tools require advanced expertise in AI and machine learning.
    \item Limited Contextualization: Few tools provide actionable insights for industry-specific or societal risks.
    \item Scalability Issues: Many tools are not optimized for seamless integration into large-scale AI workflows.
\end{itemize}

To effectively address the risks of human rights violations, future tools should try to invest more on incorporating multidisciplinary perspectives by engaging jurists, sociologists, and domain experts alongside technical teams. They should also enhance the usability of tools, simplifying interfaces, and provide detailed tutorials for non-technical stakeholders. They should also expand the scope of studies by developing integrated frameworks that assess all aspects of reliability, accountability, and transparency in a unified manner.

\section{Proposed Methodology for AI Assessment}
\label{sec:methodology}

\subsection{Overview of the Methodology}
\label{subsec:methodology-overview}

\begin{figure}[]
    \centering
    \includegraphics[width=0.83\textwidth]{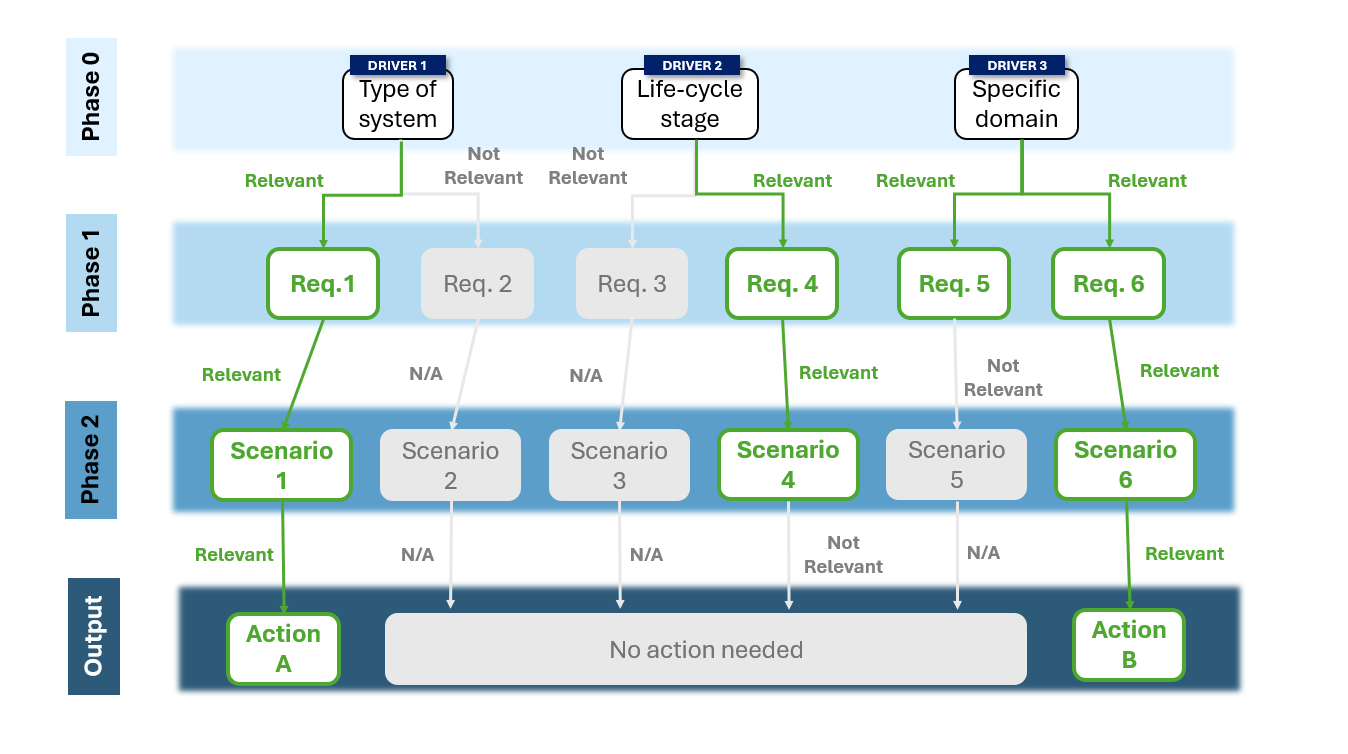}
    \caption{Overview of the FRIA Methodology: a gate-based impact assessment framework.}
    \label{fig:methodology-overview}
\end{figure}

This chapter introduces the \textit{Fundamental Rights Impact Assessment} (FRIA) methodology by HH4AI, specifically developed to assess and mitigate the potential impacts of systems on fundamental rights. The current methodology is designed for organizations seeking compliance with the AI Act while ensuring that their systems adhere to fundamental human rights principles. By employing a \emph{gate-based} structure with three main phases plus a concluding output stage (see \Cref{fig:methodology-overview}), the methodology streamlines the analysis process and ensures that only the most relevant impact progress to detailed evaluation.

At the core of the methodology is a structured assessment framework based on well-defined \textbf{impact domains} and \textbf{guiding criteria}. The impact domains cover key dimensions of AI-related impacts, including \emph{Data Governance}, \emph{Human Oversight and Control}, and \emph{Fairness and Non-Discrimination}. These guiding criteria serve as reference points for assessing AI systems’ alignment with fundamental rights and regulatory requirements.

To ensure relevance and efficiency, the methodology employs a \textbf{filtering mechanism} driven by key factors, referred to as "drivers", such as the type of system, its life cycle stage, and its domain of application. This structured filtering ensures that only applicable impacts and evaluation criteria are considered, avoiding unnecessary assessments. The Human Rights Checklist in Phase 1 serves as the primary tool for this evaluation, presenting targeted questions that assess whether an AI system’s functionalities pose impacts warranting deeper analysis. Based on the results of this phase, the methodology identifies which impacts need further examination through defined \textbf{impact scenarios}.

Impact scenarios play a crucial role in the methodology, illustrating concrete situations where an AI system could compromise fundamental rights. Each scenario undergoes a structured \textbf{self-evaluation}, assessing its relevance, severity, and the effectiveness of existing impact mitigation measures. This evaluation considers multiple dimensions, including the impact on individuals and society, the difficulty of reversing potential harm, and the duration of the consequences. Scenarios classified as relevant trigger specific remediation actions to mitigate impacts.

Building on this structured foundation, the methodology advances through three progressive phases, introduced at a high level earlier, which are described in detail in \Cref{subsec:methodology-phases}. Upon completion of the assessment, the methodology generates a comprehensive \textbf{final output}, as explained in \Cref{subsec:final-output}. This output consolidates the assessment findings in both graphical and tabular form, summarizing identified impacts, the effectiveness of existing controls, and recommended mitigation actions. In doing so, it provides decision-makers with a clear, actionable overview of the AI system’s impact, thereby facilitating effective impact management and regulatory compliance.

A key differentiator of this methodology is its \emph{gate-based} approach, ensuring efficiency by progressively refining the analysis and focusing only on the most relevant impacts. This stepwise refinement prevents unnecessary assessments, optimizes resource allocation, and enhances the clarity of impact evaluation. The methodology’s structured yet flexible design allows it to adapt to various AI applications while maintaining a rigorous human rights framework. The benefits of this approach extend beyond compliance; by embedding ethical considerations and proactive impact management into the AI life cycle, it enhances transparency, accountability, and trust in AI systems. These aspects, along with other key advantages, are explored in \Cref{subsec:innovation-benefits}, where the methodology’s innovations and benefits are analyzed in detail.

Finally, \Cref{subsec:methodology-remarks} presents concluding reflections on the methodology’s strengths, particularly its structured adaptability and role in reinforcing human rights protections throughout the AI system’s life cycle. This final discussion underscores how the methodology ensures a systematic and effective approach to human rights impact assessment, supporting both regulatory compliance and ethical AI governance.

\subsection{Phases of the Methodology}
\label{subsec:methodology-phases}
We present here a detailed explanation of each phase of the methodology, describing the key elements that compose each phase, their interactions, the specific outputs they produce, and their connection to the subsequent phase.

\subsubsection{Phase 0 - AI System Overview}
Phase 0 establishes the foundation for the impact assessment process by gathering essential information about the AI system. It defines the system’s purpose, identifies key stakeholders, and outlines the operational context. Additionally, it includes domain applicability questions to determine whether the system operates in sensitive areas, such as biometric data, processing or critical decision-making, which influence the selection of checklist questions in Phase 1. Similarly, it defines the system’s life cycle stage (e.g., development, deployment, or post-deployment), ensuring that the subsequent assessment is tailored to its current state.

Another crucial aspect of this phase is establishing a dedicated process for maintaining and updating the AI System Overview, including clear accountability for the individuals responsible. This ensures that the assessment remains accurate and reflects any changes to the system over time. By setting out these responsibilities and procedures from the outset, the output of Phase 0 provides a clear and well-defined scope for the assessment, laying the groundwork for identifying potential impacts in the following phase.

\begin{figure}[]
    \centering
    \includegraphics[width=0.8\textwidth]{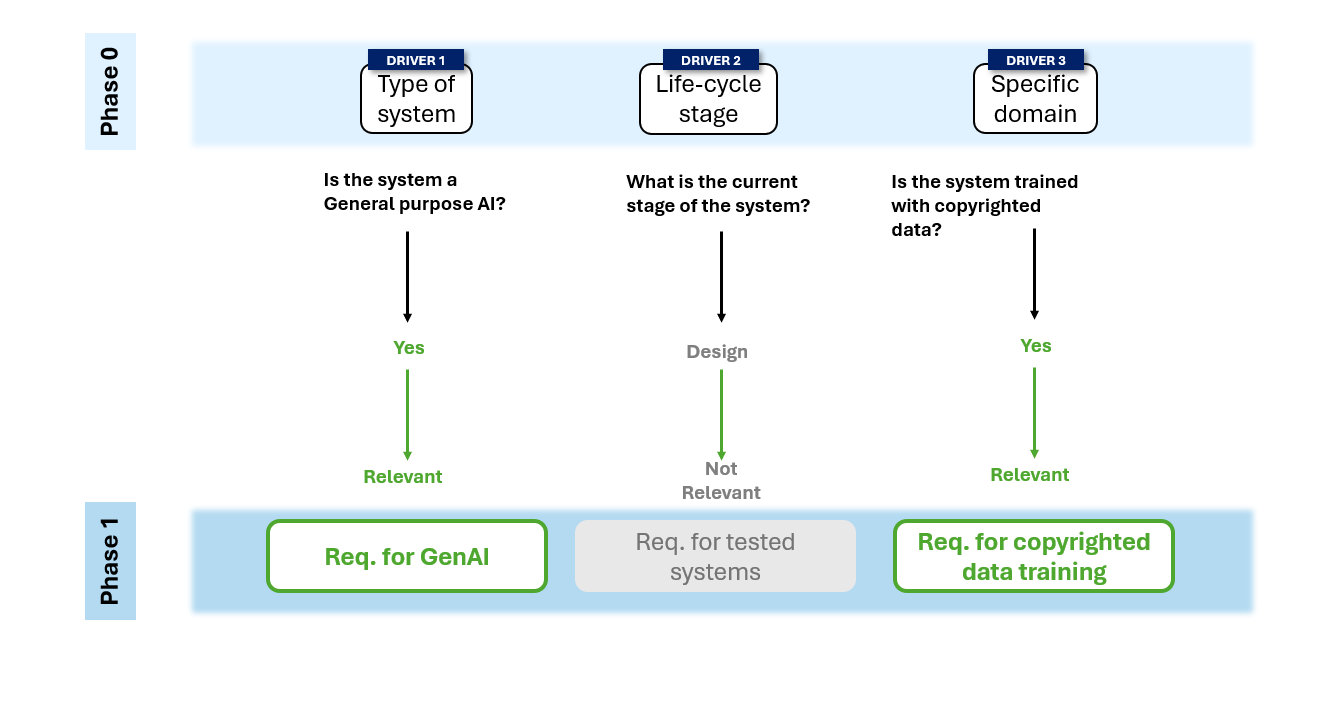}
    \caption{Transition from Phase 0 to Phase 1: identifying relevant requirements.}
    \label{fig:phase0-to-phase1}
\end{figure}

As shown in \Cref{fig:phase0-to-phase1}, the transition from Phase 0 to Phase 1 follows a structured filtering process. This ensures that only the most relevant requirements proceed for further evaluation, optimizing the efficiency of the assessment.

\subsubsection{Phase 1 - Human Rights Checklist} 
Phase 1 systematically identifies potential human rights impacts through a structured Human Rights Checklist. This checklist is designed to assess the AI system’s impact by linking each evaluation question to \textit{guiding criteria}, which are directly mapped to fundamental rights.

To ensure contextual relevance, the checklist questions are dynamically filtered based on two key factors: the system’s life cycle stage and its domain applicability. This tailored approach ensures that only questions relevant to the specific AI system under evaluation are considered. Each checklist item is also assigned to specific internal stakeholders, ensuring that subject-matter experts evaluate the areas where they have direct oversight and expertise.

\begin{figure}[]
    \centering
    \includegraphics[width=0.77\textwidth]{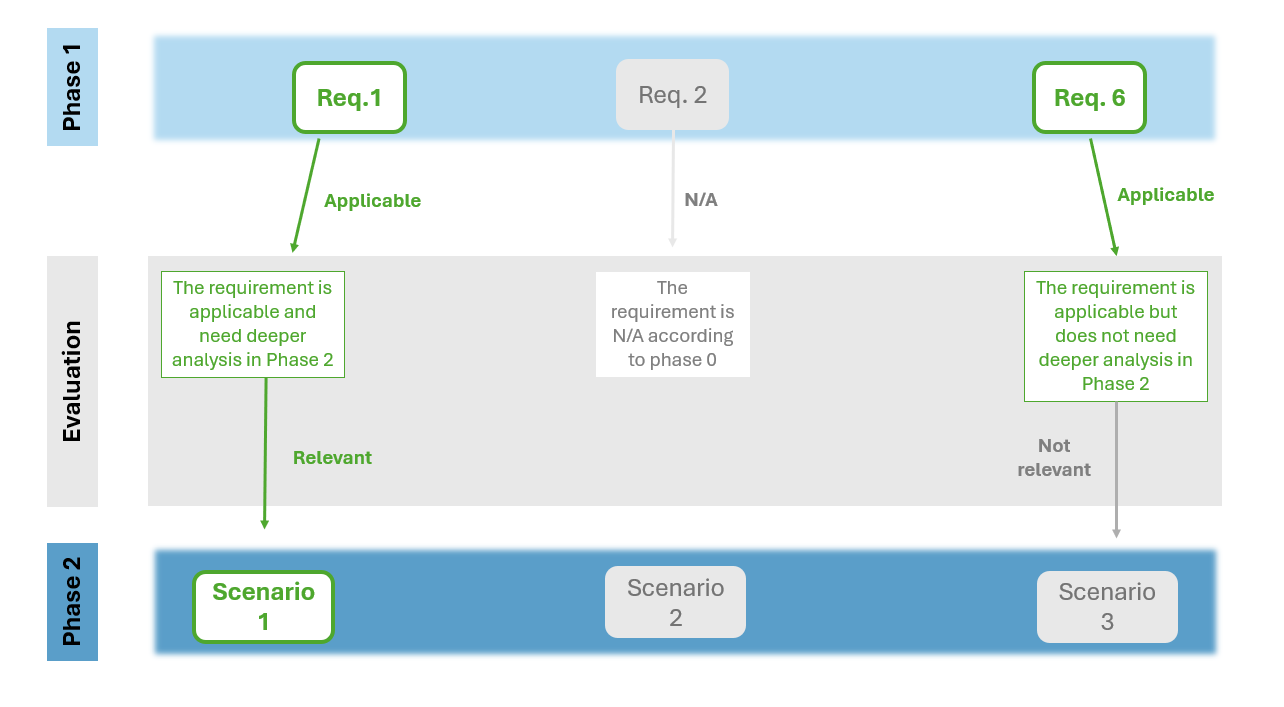}
    \caption{Transition from Phase 1 to Phase 2: identifying relevant impact scenarios.}
    \label{fig:phase1-to-phase2}
\end{figure}

The relevance of each criterion is determined through the responses to the checklist. If a criterion receives a high relevance score, indicating a potentially significant impact on fundamental rights in the context of the specific AI system under evaluation, then the assessment proceeds to Phase 2, where a more detailed analysis is conducted. This transition from Phase 1 to Phase 2 follows a structured filtering process, as illustrated in \Cref{fig:phase1-to-phase2}, ensuring that only the most critical impacts advance to deeper evaluation while optimizing efficiency.

\subsubsection{Phase 2 - Impact Assessment}
Phase 2 involves a detailed evaluation of the impacts identified in Phase 1, focusing on \textit{multiple impact scenarios} for each guiding criterion. These scenarios are designed to assess a wide range of potential impacts to fundamental rights, including ethical, legal, and social implications. The internal stakeholder responsible for each criterion conducts this assessment, determining whether effective controls exist within the organization to mitigate the identified impacts. Stakeholders are required to provide documentation or other evidence demonstrating the effectiveness of these controls, as well as to specify the individual or department responsible for maintaining and overseeing them.  

The impact assessment considers multiple evaluation dimensions to ensure a comprehensive understanding of each impact scenario. Stakeholders assess:
\begin{itemize}
    \item The \textbf{effect on individuals}, analyzing the potential impact on individual rights (e.g., privacy violations, discrimination).
    \item The \textbf{effect on society}, considering broader societal implications (e.g., increased inequality, biases in decision-making).
    \item The \textbf{effort required to mitigate or reverse the impact}, evaluating how difficult it would be to address the issue once it has occurred.
    \item The \textbf{duration of the effect}, estimating whether the impact is short-term, long-term, or potentially irreversible.
\end{itemize}

The evaluation process is structured around a three-level self-evaluation scale, where each impact scenario is classified as:
\begin{itemize}
    \item \textbf{Relevant}: the scenario poses a significant impact to fundamental rights and requires immediate action.
    \item \textbf{Partially Relevant}: the scenario presents moderate impacts that may require intervention but are not immediately critical.
    \item \textbf{Irrelevant}: the scenario does not apply or has no meaningful impact on fundamental rights.
\end{itemize} 

For each scenario assessed as Relevant or Partially Relevant, a remedial action is proposed to mitigate the identified impact. The remediation process includes:
\begin{itemize}
    \item \textbf{Action Type}: the category of intervention (e.g., policy revision, additional control implementation, training, or awareness programs).
    \item \textbf{Action Description}: a detailed explanation of the corrective measure and how it will mitigate the identified impact.
    \item \textbf{Action Owner}: The responsible individual, team, or department ensuring the implementation and effectiveness of the corrective action.
\end{itemize}

Once all impact scenarios have been evaluated and appropriate remedial actions suggested, the final classification of the impact on fundamental rights is determined for each guiding criterion. If multiple relevant impact scenarios are identified, additional mitigation strategies may be necessary to ensure compliance and impact reduction. However, if most scenarios are classified as Irrelevant, no further action or in-depth analysis is required for that specific criterion.  

This structured, multi-dimensional approach ensures that AI-related impacts to fundamental rights are systematically identified, assessed, and mitigated, while maintaining accountability and transparency throughout the process.  

\begin{figure}[]
    \centering
    \includegraphics[width=0.8\textwidth]{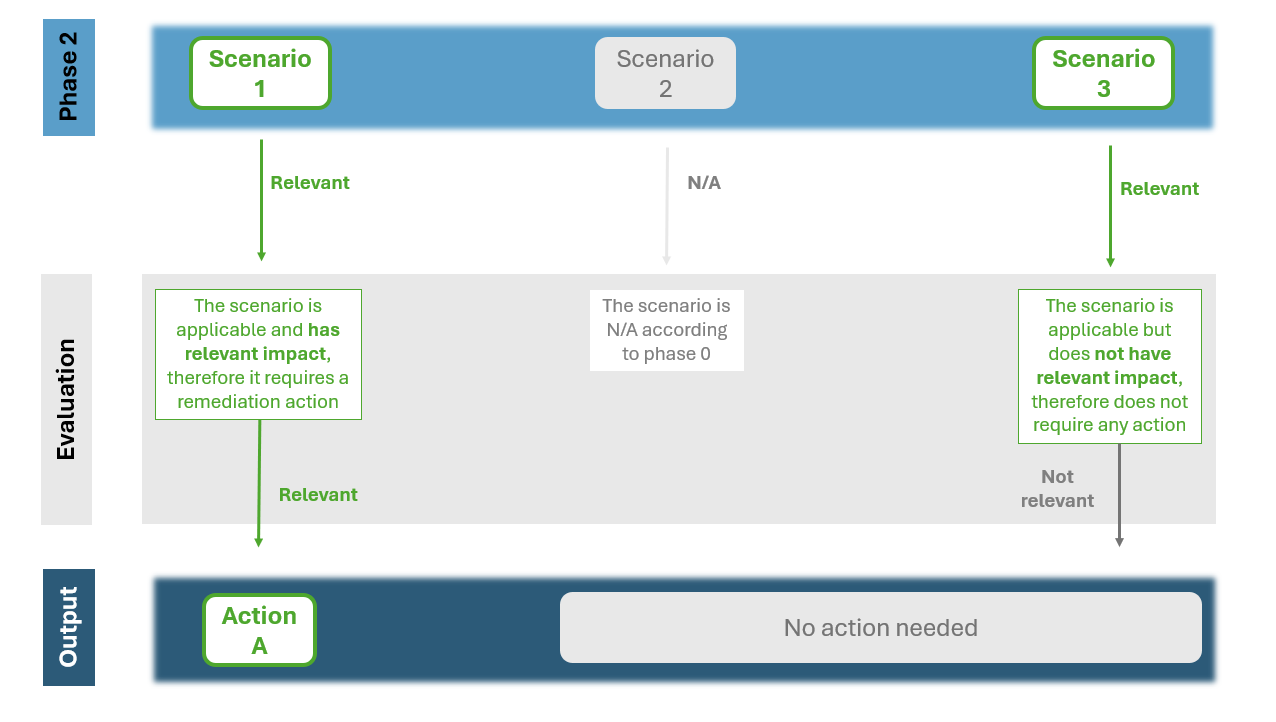}
    \caption{Transition from Phase 2 to Output: identifying required remediation actions.}
    \label{fig:phase2-to-output}
\end{figure}

As illustrated in \Cref{fig:phase2-to-output}, the transition from Phase 2 to the Output stage ensures that only scenarios classified as relevant and having a significant impact require corrective actions. If a scenario is deemed relevant but without a significant impact, no further action is required. Scenarios classified as not relevant are excluded from the final output. This structured filtering approach ensures that remediation efforts are targeted, efficient, and aligned with the identified impacts, maintaining an effective and accountable impact assessment process.

\subsection{Final Output}
\label{subsec:final-output}
The final output of the Fundamental Rights Impact Assessment provides a comprehensive summary of the assessment results across all phases. This output consists of both graphical and tabular representations to facilitate a clear and structured interpretation of the evaluation process.  

The tabular overview presents a structured breakdown of the assessments and evaluations conducted in Phase 1 and Phase 2, detailing relevance scores, stakeholder responses, and identified impacts. The graphical overview complements this by offering a visual representation of key insights, ensuring an intuitive and easily digestible format for decision-makers.  

The final output is structured into two primary components:
\begin{itemize}
    \item An overview of results, which includes both the graphical and tabular representations of the assessment conducted in Phase 1 (requirements analysis) and Phase 2 (impact scenario evaluation).
    \item A remediation actions section, detailing the list of required actions, their types, and the responsible stakeholders for implementation.
\end{itemize}

The final output ensures that all identified impacts and corresponding remediation actions are documented in a structured manner. The graphical and tabular overviews provide a clear impact profile, while the remediation section ensures accountability by assigning ownership to corrective actions. This comprehensive output enables decision-makers to track, evaluate, and implement impact mitigation strategies effectively, ensuring that fundamental rights considerations are addressed throughout the AI system’s life cycle.

\subsection{Innovation and Benefits}
\label{subsec:innovation-benefits}
The FRIA methodology introduces several key innovations and benefits, enhancing the effectiveness and applicability of AI impact assessments while ensuring a structured and actionable approach to impact mitigation.

\begin{itemize}
 \item \textbf{Detailed impact Scenario Analysis}: by defining multiple scenarios for each guiding criterion, the methodology enables a comprehensive evaluation of potential impacts. This granular approach ensures a thorough understanding of how an AI system may impact fundamental rights and allows for the development of precise, targeted mitigation strategies.
 \item \textbf{Stakeholder-Driven Evaluation}: the assessment process integrates the expertise of internal stakeholders, leveraging their real-world insights into system design, deployment, and governance. This ensures that impact identification and mitigation strategies are based on practical knowledge of existing controls and operational impacts.
 \item \textbf{Self-Evaluation Scale}: a standardized three-level scale (Relevant, Partially Relevant, or Irrelevant) quantifies the significance of each identified impact. This structured approach facilitates clear decision-making and ensures that only substantial impacts advance to deeper analysis and remediation.
 \item \textbf{Human Rights Mapping}: impacts and scenarios are systematically categorized based on guiding criteria linked to fundamental rights. This structured alignment provides organizations with a transparent, legally grounded understanding of how AI functionalities may affect individual rights.
 \item \textbf{Flexibility and Context-Specific Adaptation}: the methodology adapts to different AI use cases by tailoring the assessment based on the system’s domain and life cycle stage. This ensures that organizations focus on relevant impacts without performing unnecessary evaluations.
 \item \textbf{Proactive impact Mitigation}: beyond identifying impacts, the methodology prescribes concrete remedial actions for scenarios deemed Relevant or Partially Relevant. These interventions, ranging from policy revisions to technical controls and training programs, ensure that the assessment process is solution-oriented, actively supporting organizations in enhancing compliance and minimizing potential harm.
\end{itemize}

\subsection{Final Remarks}
\label{subsec:methodology-remarks}
The FRIA methodology provides a structured, systematic, and scalable framework for assessing and mitigating the impact of AI systems on fundamental rights. By following a gate-based approach, it ensures that only the most relevant impacts undergo detailed evaluation, optimizing resources while maintaining a high level of scrutiny. This structured assessment process enables organizations to integrate ethical considerations, regulatory compliance, and impact management into AI development and deployment strategies.  

The methodology not only identifies and evaluates impacts but also assesses the effectiveness of existing safeguards and establishes accountability for their continuous monitoring. The final output offers a comprehensive overview of impact levels and required remediation actions, ensuring that decision-makers have a clear understanding of potential impacts and the necessary steps to mitigate them. This structured approach enhances transparency in AI governance, making impact assessment results both accessible and actionable.  

Beyond regulatory compliance, the methodology fosters a proactive approach to responsible AI development by embedding fundamental rights considerations throughout the AI system life cycle. This allows organizations to move beyond a reactive compliance mindset toward continuous improvement in AI ethics and governance. The structured remediation process ensures that identified impacts are not only acknowledged but also addressed through concrete actions, reinforcing accountability and fostering trust in AI systems.  

By systematically aligning AI impact assessment with human rights principles and governance best practices, the HH4AI FRIA methodology supports organizations in achieving AI accountability, regulatory alignment, and ethical governance. It provides a robust framework for mitigating AI-related impacts while promoting sustainable and responsible AI development, ensuring that fundamental rights remain a priority in the design, deployment, and operation of AI systems.

\section{Case Study: Automated Triage Service in Health Care}
\label{sec:case-study}
To illustrate the application of the \textit{Fundamental Rights Impact Assessment} (FRIA) methodology, this section presents a toy example of an \textbf{Automated Triage Service in Health Care}. The system assists medical personnel by gathering preliminary patient information, then producing structured reports to support clinical decision-making. It leverages \textit{retrieval-augmented generation} (RAG) techniques and large language models (LLMs) to handle patient inquiries, extract relevant medical and administrative details and generate outputs for healthcare professionals. Operating in a healthcare setting, the system raises substantial concerns regarding data protection, fairness and responsible oversight—making it a suitable test bed for evaluating the FRIA methodology.

The analysis focuses on three key \textit{guiding criteria}: \textbf{Data Governance}, \textbf{Human Oversight and Control}, and \textbf{Fairness \& Non-Discrimination}. These criteria cover core dimensions of AI-driven decision support in healthcare, emphasizing data protection, accountability and equitable treatment.

\subsection{Phase 0: AI System Overview}
Phase 0 defines the scope and context of the AI system under assessment. Here, the \textbf{Automated Triage Service in Health Care} is identified as a chatbot-based application designed to facilitate triage by collecting patient information, processing medical visit requests and generating structured reports. Because it relies on personal data and exerts direct influence on clinical decision-making, particular attention is paid to human oversight and compliance requirements.

This phase also considers governance structures and life cycle stages, enabling alignment with relevant regulatory and ethical guidelines. The insights gathered here inform the filtering process that takes place in Phase 1.

Since the system does not rely on copyrighted data or generative AI functionalities, and is currently in the implementation phase, related requirements (e.g., advanced generative features, monitoring during the use phase) are excluded from further analysis.

\subsection{Phase 1: Human Rights Checklist}
Phase 1 employs a \textit{Human Rights Checklist} to filter potential impact areas. Each question in the checklist is linked to a guiding criterion mapped to fundamental rights. This structured assessment determines whether an issue warrants deeper exploration in Phase 2.

\begin{figure}[]
    \centering
    \includegraphics[width=\linewidth]{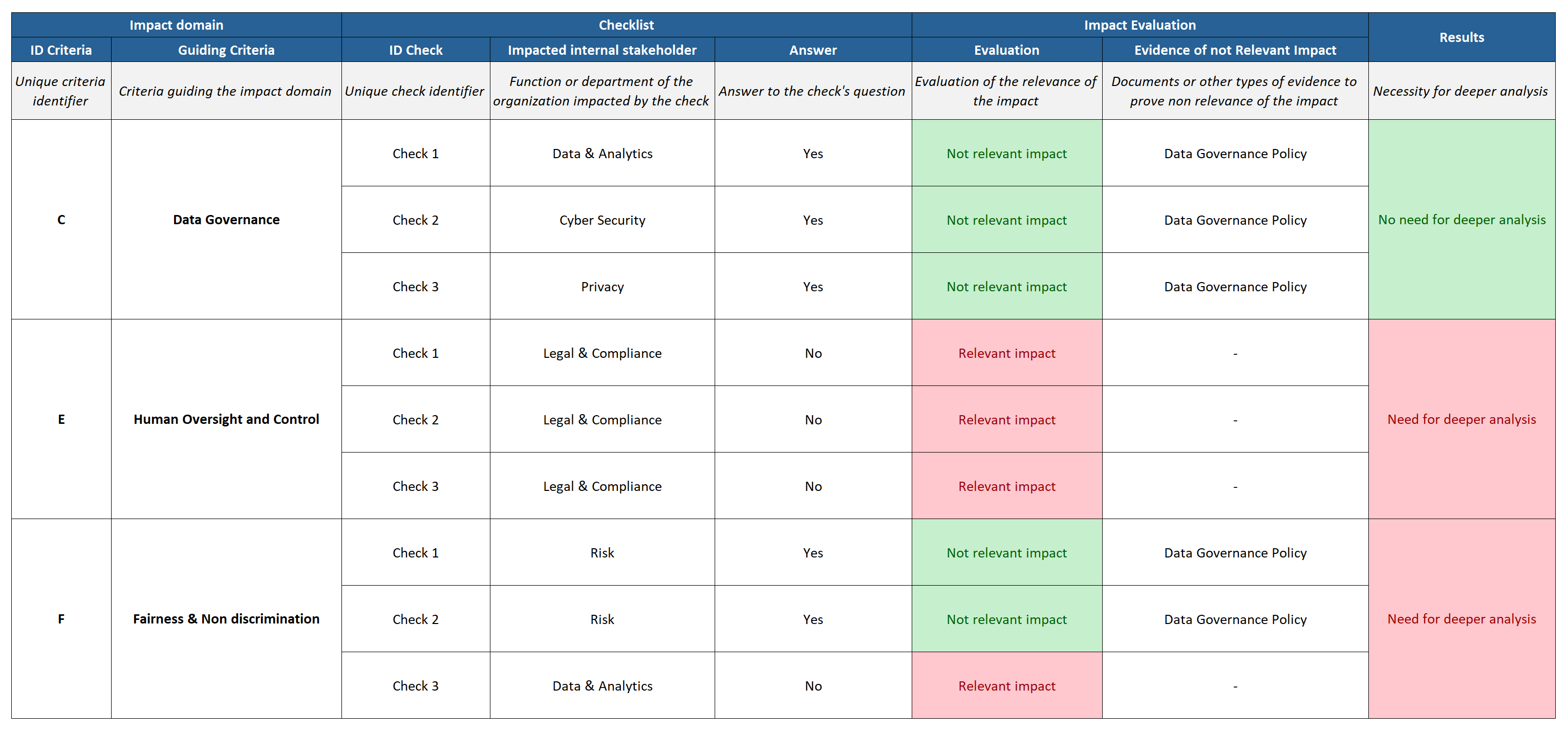}
    \caption{Phase 1 checklist - Case Study}
    \label{fig:phase1_toy}
\end{figure}

From the checklist, three significant findings (as shown in \Cref{fig:phase1_toy}) emerged:

\begin{enumerate}
\item \textbf{Data Governance\\} Established policies minimize personal data processing and ensure traceability. Verification mechanisms for external data sources are in place, raising no immediate compliance concerns about AI-generated recommendations. No further action is needed for this domain in Phase 2.
\item \textbf{Human Oversight and Control\\} The absence of formalized human review and override mechanisms for AI-generated recommendations presents a risk of automation bias. This gap requires deeper investigation in Phase 2.
\item \textbf{Fairness \& Non-Discrimination\\} Although the system relies on static decision rules rather than adaptive learning, limited accessibility features risk excluding certain user demographics. The need to address potential bias and improve inclusivity makes this domain relevant for Phase 2.
\end{enumerate}

Based on these results, the primary focus of Phase 2 is directed toward improving oversight mechanisms and addressing fairness concerns, including accessibility enhancements.

\subsection{Phase 2: Impact Assessment}
Phase 2 evaluates specific \textbf{impact scenarios} within each criterion identified as warranting further analysis. Each scenario is assessed by the responsible stakeholder, considering four dimensions: effects on individuals, effects on society, the effort required for mitigation and the duration of potential consequences.

A three-level self-evaluation scale (\textit{Relevant, Partially Relevant} or \textit{Irrelevant}) synthesizes these considerations to prioritize remediation measures effectively.

As no concerns emerged under \textbf{Data Governance} in Phase 1, no scenario is analyzed for that criterion. By contrast, the following areas were flagged:

\begin{itemize}
    \item \textbf{Human Oversight and Control\\} AI-generated recommendations were being accepted without a structured validation process, introducing automation bias. This scenario was deemed \textit{Relevant}, prompting the introduction of \textit{decision override protocols} and \textit{performance monitoring guidelines}, which ultimately reduced the remaining impact to \textit{Partially Relevant}.
    \item \textbf{Fairness \& Non-Discrimination\\} Underrepresentation of certain demographic groups necessitated measures such as \textit{bias assessment documentation} and \textit{inclusivity compliance evaluations}, classifying this scenario as \textit{Relevant}.
\end{itemize}

\begin{figure}[]
    \centering
    \includegraphics[width=\linewidth]{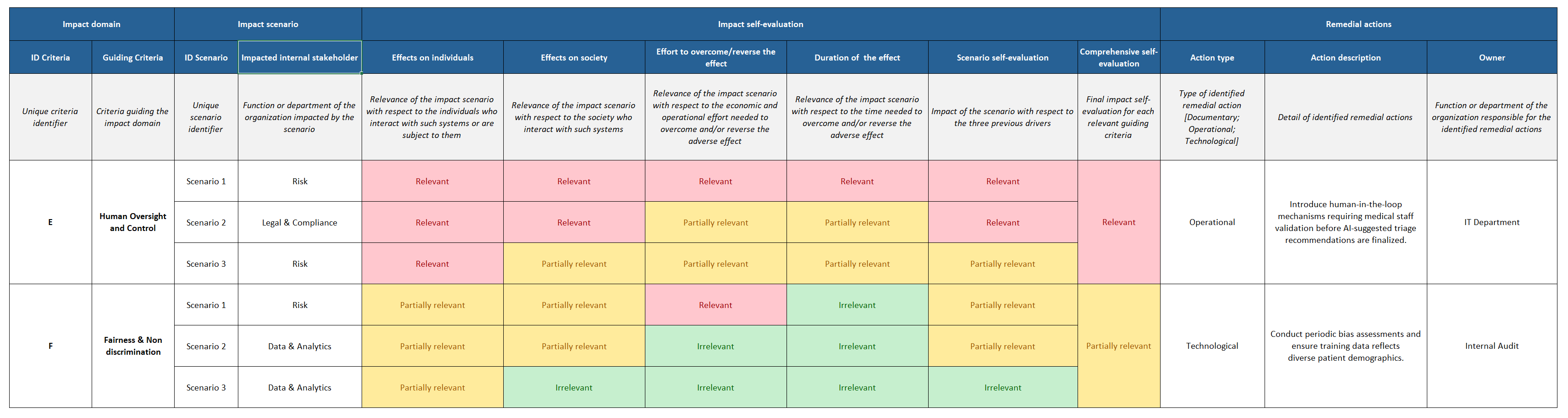}
    \caption{Phase 2 Self Evaluation - Case Study}
    \label{fig:phase2_toy}
\end{figure}

These findings and their evaluations are summarized in \Cref{fig:phase2_toy}.

\subsection{Final Output: Mitigation and Remedial Actions}
\begin{figure}[]
    \centering
    \includegraphics[width=\linewidth]{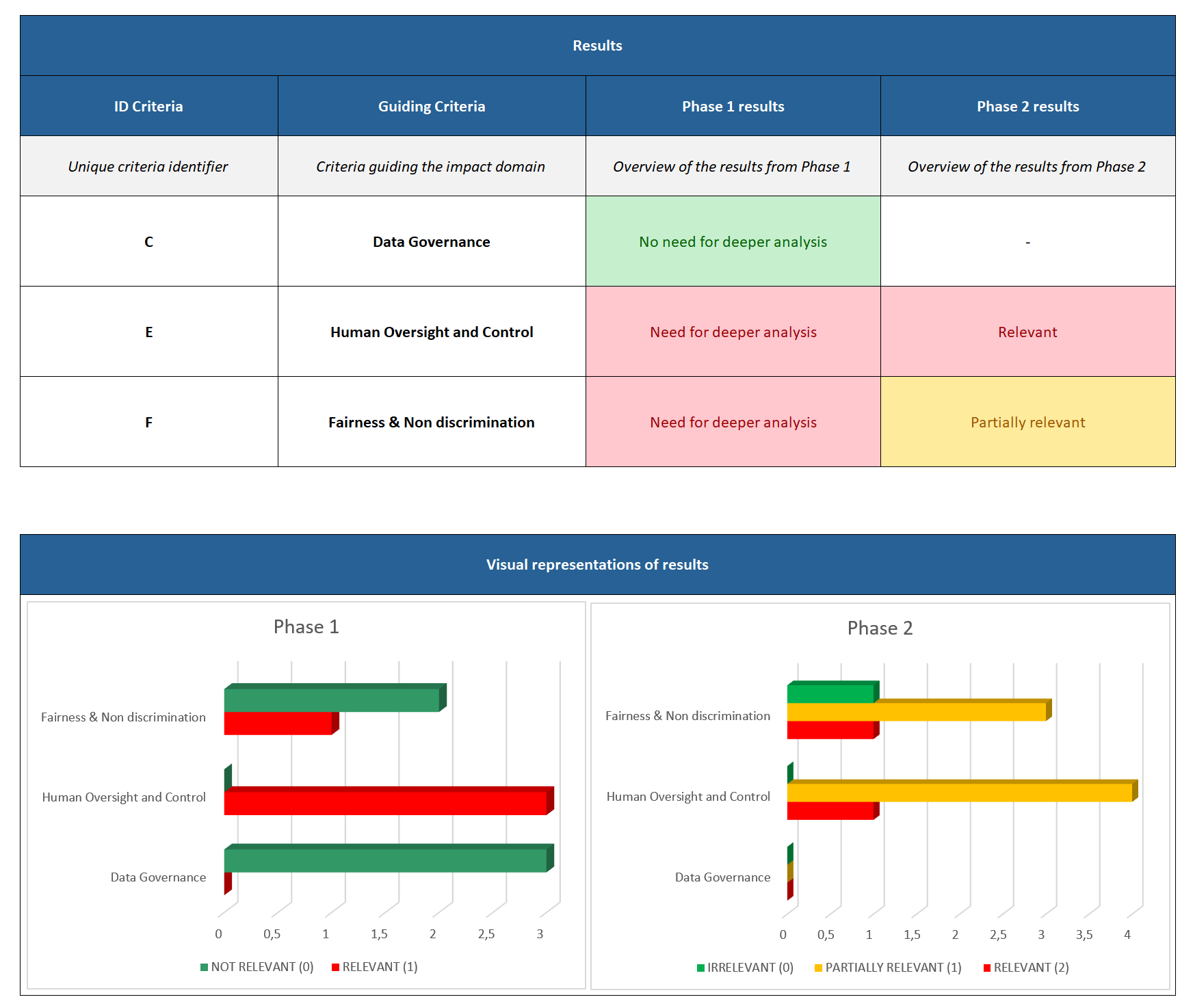}
    \caption{Output - Case Study}
    \label{fig:output_toy}
\end{figure}

The final output consolidates Phase 2 findings into a practical plan for addressing identified risks (see \Cref{fig:output_toy}). Although the actions described here are illustrative, they demonstrate how the FRIA methodology translates insights from self-evaluation into concrete solutions:
\begin{itemize}
    \item \textbf{Human Oversight and Control Remediation\\} Introduce \textit{human-in-the-loop mechanisms} requiring medical staff approval before finalizing triage recommendations.
    \item \textbf{Fairness \& Non-Discrimination Remediation\\} Conduct \textit{periodic bias assessments} and revise training data to ensure broader demographic coverage.
\end{itemize}

\subsection{Key Takeaways}
In summary, this toy example highlights how systematic filtering, scenario-based impact assessment, and targeted mitigation measures can integrate fundamental rights considerations throughout the AI life cycle. While the proposed interventions are illustrative, they showcase the FRIA methodology’s capacity to focus on the most pressing vulnerabilities (those classified as \textit{Relevant} or \textit{Partially Relevant}) and provide a pathway for organizational action. Through this approach, the methodology aims to enhance compliance, minimize undue harm and align AI systems more closely with human rights principles.

\section{Discussion and Future Work}
\label{sec:discussion}
The gate-based framework presented in this paper offers a focused method for identifying and mitigating AI systems’ impacts on fundamental rights. Its phased structure, coupled with a tailored filtering mechanism, channels organizational resources toward the most critical risks, thereby reducing superfluous evaluations. This approach not only supports compliance with emerging regulations but also promotes transparency and accountability across varied AI life cycles.

A notable advantage of this design is its balance between flexibility and rigor. The methodology caters to a wide spectrum of AI applications, ensuring that essential concerns—such as accountability, literacy and data governance—are appropriately addressed. Simultaneously, detailed \emph{impact scenarios} allow for deeper scrutiny of higher-risk functionalities, preventing resource misallocation and clarifying subsequent remediation activities.

However, certain challenges remain. Many industries require domain-specific adaptations to address diverse regulatory environments, while the rapid evolution of AI technology demands periodic revisions to both the checklist items and the overarching guidance framework. Moreover, effective remediation hinges on the maturity of each organization, the availability of specialized personnel, and robust governance infrastructures. 

Looking ahead, an important area of future work is the enhancement of Phase 2 through quantitative metrics. In particular, measuring dimensions such as fairness, reliability and system transparency with numerical indicators can sharpen risk estimation and enable more robust benchmarking across diverse AI systems. This data-driven perspective complements the qualitative assessments, helping to prioritize remediation efforts and ensuring that any identified gaps or vulnerabilities are addressed through targeted, evidence-based strategies.

\section{Conclusion}
\label{sec:conclusion}
The work presented in this paper responds to the growing need for a structured, comprehensive methodology to evaluate the impact of AI systems on fundamental rights. By adopting a gate-based structure with progressive filtering, the proposed framework directs focused attention to potentially critical issues, while reducing unnecessary assessments and promoting efficient resource allocation. Its flexible design accommodates diverse AI application domains and life cycle stages, supporting both high-level governance requirements and in-depth scenario analysis.

In practice, the methodology enables organizations to identify key areas of risk, evaluate existing controls and propose concrete mitigation strategies. The phased approach consolidates qualitative insights from checklists and scenario-based evaluations and will soon incorporate quantitative metrics for finer-grained assessments. This evolution, combined with planned engagements involving external stakeholders, reinforces the methodology’s adaptability, rigor and relevance in real-world contexts.

Overall, the presented framework contributes a scalable and actionable solution for guiding organizations toward responsible AI development and deployment, aligning technical measures with ethical principles and regulatory demands. Its implementation aims to foster greater transparency, accountability and trust, underscoring the critical role of human rights considerations at every stage of the AI system life cycle.

\section*{Acknowledgments}
The work reported in this paper has been partly funded by the European Union - NextGenerationEU, under the National Recovery and Resilience Plan (NRRP) Mission 4 Component 2
Investment Line 1.5: Strengthening of research structures and creation of R\&D ``innovation
ecosystems'', set up of ``territorial leaders in R\&D'', within the project ``MUSA - Multilayered
Urban Sustainability Action'' (contract n. ECS 00000037).

\newpage
\thispagestyle{empty}
\bibliographystyle{unsrt}  
\bibliography{templateArxiv}

\end{document}